\documentclass[conference]{IEEEtran}
\IEEEoverridecommandlockouts
\usepackage{cite}
\usepackage{amsmath,amssymb,amsfonts}
\usepackage{algorithmic}
\usepackage{graphicx}
\usepackage{textcomp}
\usepackage{xcolor}
\usepackage{subfigure}
\usepackage{multirow}
\def\BibTeX{{\rm B\kern-.05em{\sc i\kern-.025em b}\kern-.08em
    T\kern-.1667em\lower.7ex\hbox{E}\kern-.125emX}}
\begin{document}

\title{HEAM: High-Efficiency Approximate Multiplier Optimization for Deep Neural Networks  \\
\thanks{This work is supported by the National Natural Science Foundation of China under grants 61971143 and 62174035. }
}

\author{\IEEEauthorblockN{Su Zheng$^*$, Zhen Li$^*$, Yao Lu, Jingbo Gao, Jide Zhang, Lingli Wang}
\IEEEauthorblockA{\textit{State Key Laboratory of ASIC \& System, \textit{Fudan University}} \\
Shanghai, China \\
\{szheng19, lizhen19, yaolu20, jbgao19, jdzhang17, llwang\}@fudan.edu.cn}
$^*$ \textit{These authors contributed equally to this work}. 
}


\maketitle

\begin{abstract}

We propose an optimization method for the automatic design of approximate multipliers, which minimizes the average error according to the operand distributions. Our multiplier achieves up to 50.24\% higher accuracy than the best reproduced approximate multiplier in DNNs, with 15.76\% smaller area, 25.05\% less power consumption, and 3.50\% shorter delay. Compared with an exact multiplier, our multiplier reduces the area, power consumption, and delay by 44.94\%, 47.63\%, and 16.78\%, respectively, with negligible accuracy losses. The tested DNN accelerator modules with our multiplier obtain up to 18.70\% smaller area and 9.99\% less power consumption than the original modules.


\end{abstract}


\section{Introduction} \label{intro}
Deep Neural Networks (DNNs) achieve tremendous success over recent years in artificial intelligence applications. Due to the large quantities of parameters, computing DNNs requires huge numbers of multiplication operations. In neural network accelerators, approximate multipliers are promising substitutes for exact multipliers, which can make a tradeoff between precision and hardware cost. To take advantage of this feature, we explore approximate multiplier design methodology to reduce the hardware costs of DNN accelerators. 

Various works show that approximate multipliers can effectively reduce the area and power consumption with small precision loss. Most approximate multipliers are designed by adopting novel approximate blocks. In \cite{Approximate:KMap}, the Karnaugh map of an exact 2$\times$2 multiplier is modified to form a basic approximate block. Larger multipliers are constructed by stacking 2$\times$2 blocks. Reference \cite{Approximate:Configurable} designs an approximate multiplier by simplifying the partial product accumulation block with limited carry propagation. Mathematical approximation methods are used to design approximate multipliers as well. Mitchell approximation is adopted to design an iterative logarithmic multiplier \cite{Approximate:Mitchell}. In \cite{Approximate:HybridRadix}, an approximate hybrid radix encoding is proposed for the generation of approximate multipliers. 

Recent studies have explored approximate computing for DNNs, such as \cite{Approximate:ANN}, \cite{Approximate:ALWANN}, and \cite{Approximate:Optimal}. An approximate multiplier generated by Cartesian Genetic Programming (CGP) is applied to neural networks in \cite{Approximate:ANN}. Multiplier-less Artificial Neuron (MAN) \cite{Approximate:Alphabet} uses a novel Alphabet Set Multiplier which utilizes a pre-computer bank and an alphabet selection procedure to reduce the computational cost and accuracy loss in neural networks. Reference \cite{Approximate:Optimal} designs an optimization scheme to build a floating-point approximate multiplier, which is optimal with the given bases. 

The approximate multipliers for DNNs above rely on optimization procedures to ensure low accuracy loss. These optimization methods have an implicit assumption that the operands are uniformly distributed in the given space. Nevertheless, according to studies on the weight analysis of neural networks such as \cite{DNN:WeightAnalysis}, the weight distributions are typically not uniform. Similarly, our experiment shows that the weights of quantized DNNs are concentrated around certain points. Reference \cite{Approximate:CGPProb} considers the weight distributions of DNNs and optimizes an approximate multiplier with CGP. However, the probability distributions of the activations are ignored, which leads to the biased estimation of the DNN computation. In the proposed method, the probability distributions of both operands are utilized to reduce the errors of multiplications on frequently-occurring operands. Furthermore, the CGP-based and MAN-multiplier-based methods require complicated DNN retraining to ensure low accuracy loss, while the proposed method does not need retraining. In addition, a precision controller is designed in \cite{Approximate:CNNInference} to obtain low accuracy loss. However, it limits the reduction of hardware cost. 

We propose an application-specific optimization method which can generate a \textbf{H}igh-\textbf{E}fficiency \textbf{A}pproximate \textbf{M}ultiplier (\textbf{HEAM}). We design an approximate multiplier for DNNs with the proposed method, achieving small area, high power efficiency, short delay, and negligible accuracy loss.

The rest of this paper is organized as follows. In Section \ref{Sec:Method}, we describe our probability-based optimization method and the optimized approximate multiplier for DNNs. In Section \ref{Sec:Exp}, we compare various multipliers in terms of hardware cost, evaluate the accuracies of the DNNs with the multipliers, and show the comparison of the DNN accelerator modules integrated with the multipliers. The last section is the conclusion. 

\section{Approximate Multiplier Optimization Method} \label{Sec:Method}


\subsection{Problem Formulation} \label{problem}

An approximate multiplication can be decomposed into the weighted sum of the outputs of several basic operations. We use $\mathbf{\theta} = [\theta_0, \theta_1, ..., \theta_{K-1}]$ to denote a weight vector with $K$ scalar elements. The approximate multiplication can be formulated as:
\begin{equation}
\label{Eqn:Formulation}
f(x, y \vert \mathbf{\theta}) = \sum_{i=0}^{K-1} \theta_i L_i(x, y)
\end{equation} 
where $L_i(x, y)$ is the output of a basic operation. We elaborate on the construction of $L_i(x, y)$ in Session \ref{SubSec:Structure}. For operands $x$ and $y$, we follow \cite{Approximate:Optimal} to define the error of $f(x, y \vert \mathbf{\theta})$ as: 
\begin{equation}
\label{Eqn:Distance}
D(x, y \vert \mathbf{\theta}) = ( xy - f(x, y \vert \mathbf{\theta}) )^2
\end{equation} 

In applications that are based on quantized numbers, $x$ and $y$ are discrete variables with finite values. Therefore, the average error of an integer approximate multiplier can be defined as: 
\begin{equation}
\label{Eqn:FixedPointError}
E_d(X_d, Y_d, \mathbf{\theta}) = \sum_{i=0}^{N-1} \sum_{j=0}^{M-1} D(x_i, y_j \vert \mathbf{\theta})p(x_i, y_j)
\end{equation} 
where $x \in X_d$, $y \in Y_d$, $X_d = \{x_0, x_1, ..., x_{N-1}\}$, and $Y_d = \{y_0, y_1, ..., y_{M-1}\}$. The constants $N$ and $M$ are the numbers of possible values of $x$ and $y$, respectively. Notation $p(x_i, y_j)$ represents the probability that $x = x_i$ and $y = y_j$. $E_d(X_d, Y_d, \mathbf{\theta})$ is the expectation of errors in the applications where $x$ and $y$ are subject to $p(x, y)$. We can reduce the computational errors by minimizing $E_d(X_d, Y_d, \mathbf{\theta})$. 



\subsection{Optimization Method Details} \label{SubSec:Structure}

\begin{figure}[tbp]
\centering
\subfigure[A 4$\times$4 exact multiplier.]{
\label{Fig:opti_method_a}
\begin{minipage}[t]{0.45\linewidth}
\centering
\includegraphics[width=0.7\linewidth]{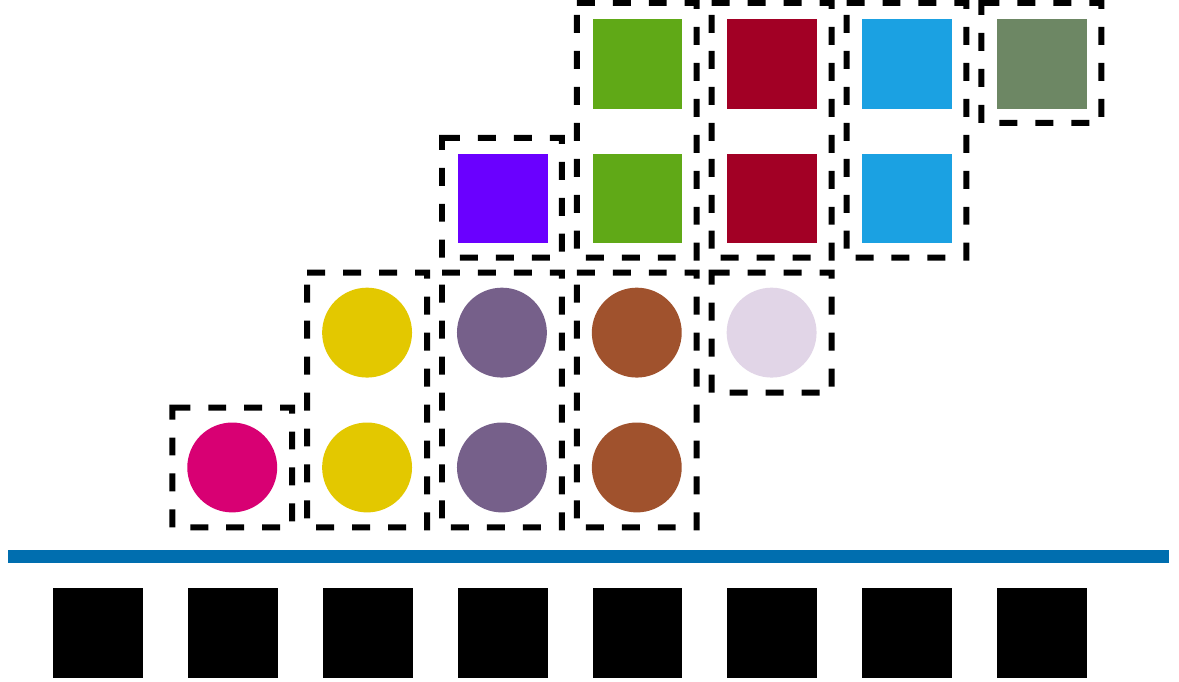}
\end{minipage}
}%
\subfigure[A 4$\times$4 approximate multiplier with compressed terms.]{
\label{Fig:opti_method_b}
\begin{minipage}[t]{0.45\linewidth}
\centering
\includegraphics[width=0.7\linewidth]{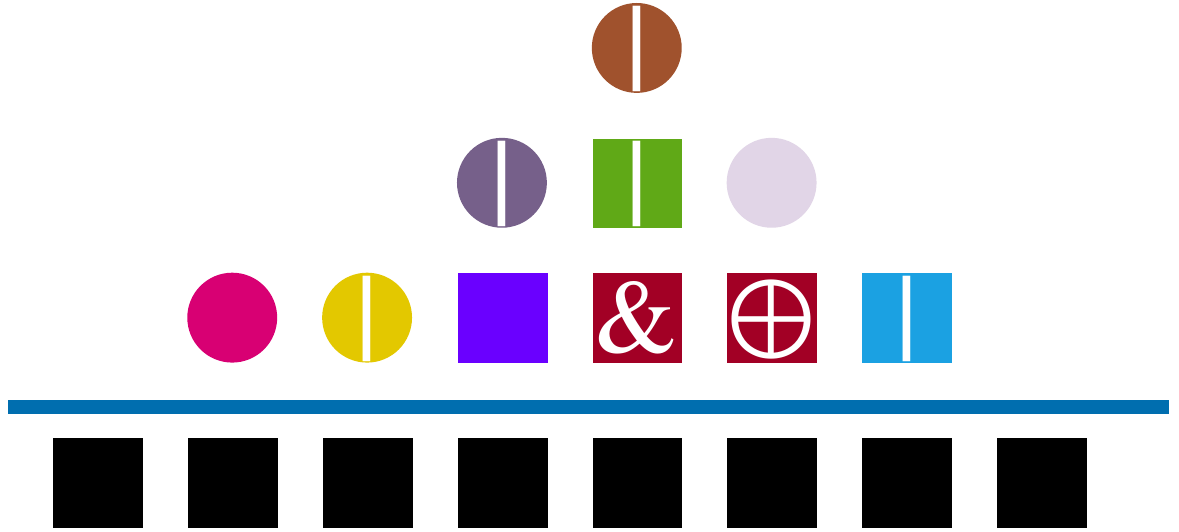}
\end{minipage}
}
\caption{An example of compressing the partial products of an 4$\times$4 multiplier.}
\label{Fig:opti_method}
\end{figure}

\begin{figure*}[tbp]
\centering
\begin{minipage}[t]{0.6\linewidth}
\subfigure[Partial products of an 8$\times$8 multiplier.]{
\label{Fig:partial_products}
\begin{minipage}[t]{0.5\linewidth}
\centering
\includegraphics[width=\linewidth]{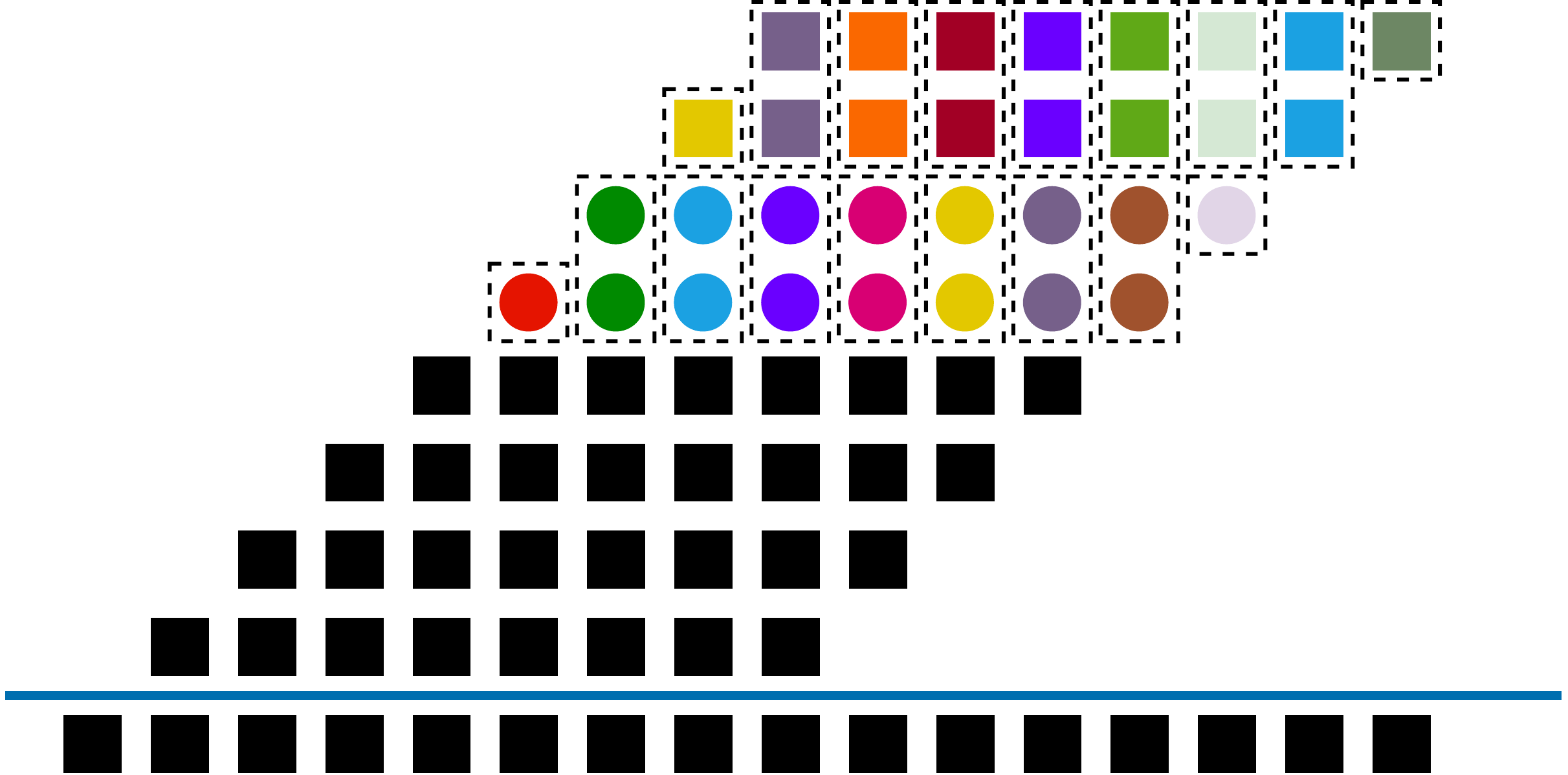}
\end{minipage}
}%
\subfigure[Multiplier with the compressed terms.]{
\label{Fig:appro_mul}
\begin{minipage}[t]{0.5\linewidth}
\centering
\includegraphics[width=\linewidth]{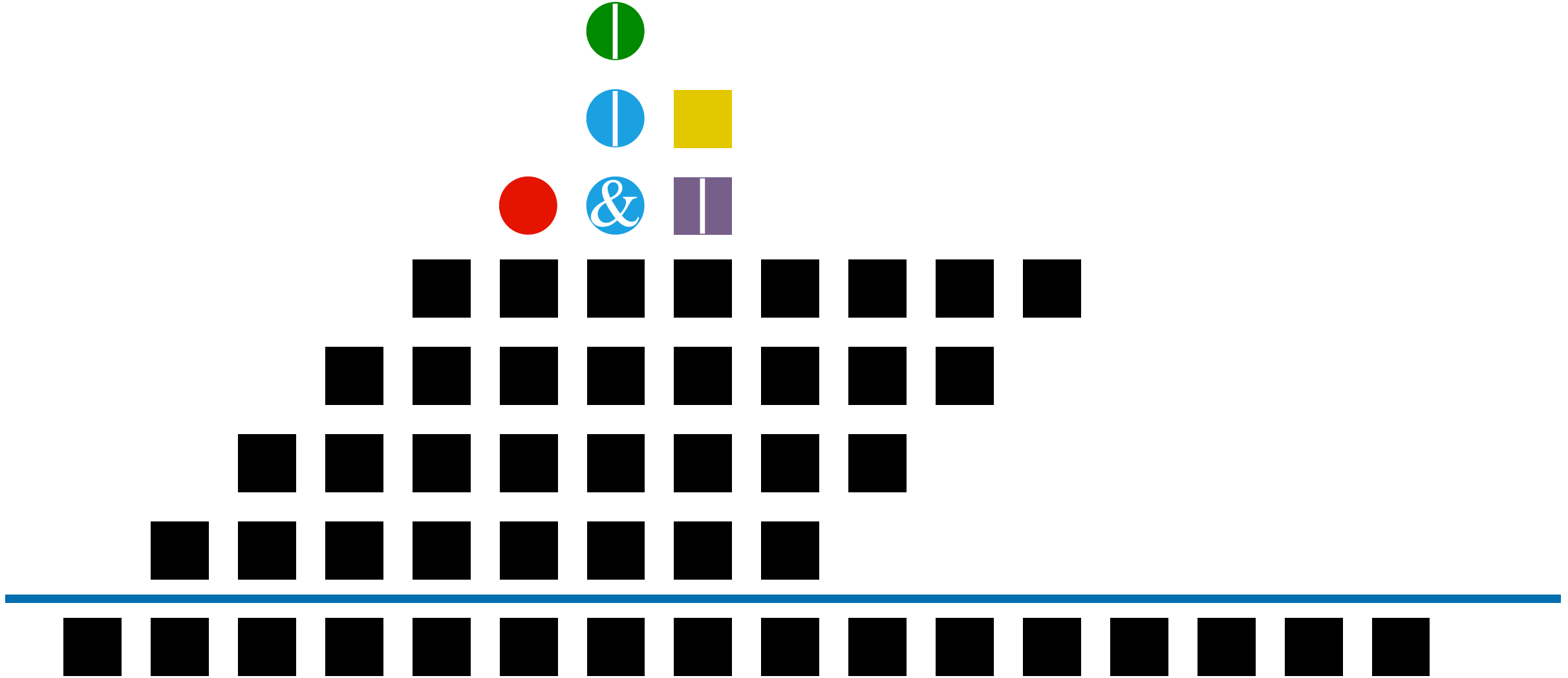}
\end{minipage}
}
\caption{The exact and approximate unsigned 8$\times$8 multipliers. }
\label{Fig:progress}
\end{minipage}%
\begin{minipage}[t]{0.4\linewidth}
\subfigure[Histogram of inputs. ]{
\label{Fig:FC1_data}
\begin{minipage}[t]{0.5\linewidth}
\centering
\includegraphics[width=\linewidth]{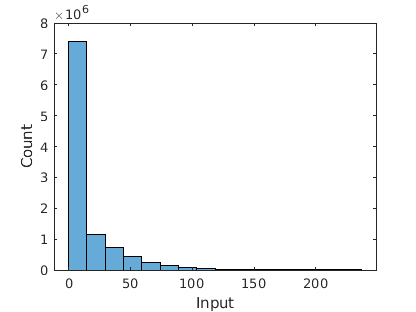}
\end{minipage}
}%
\subfigure[Histogram of weights. ]{
\label{Fig:FC1_weight}
\begin{minipage}[t]{0.5\linewidth}
\centering
\includegraphics[width=\linewidth]{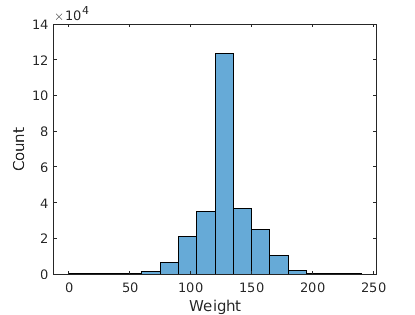}
\end{minipage}
}
\caption{The histograms of the quantized inputs and weights of the \textit{FC1} layer of LeNet on MNIST dataset. }
\label{Fig:FC1_distribution}
\end{minipage}
\centering
\end{figure*}

\begin{figure*}[tbp]
\subfigure[Area comparison.]{
\label{Fig:compare_7nm_area}
\begin{minipage}[t]{0.45\linewidth}
\centering
\includegraphics[width=0.9\linewidth]{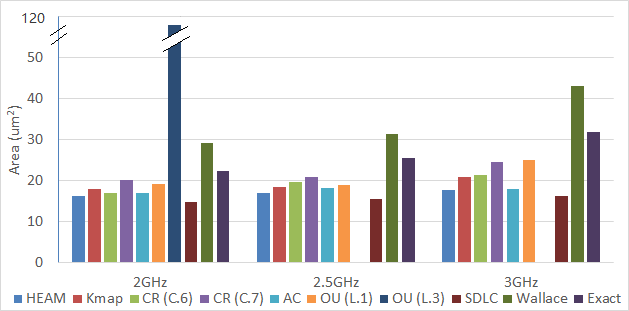}
\end{minipage}
}%
\subfigure[Power consumption comparison.]{
\label{Fig:compare_7nm_power}
\begin{minipage}[t]{0.45\linewidth}
\centering
\includegraphics[width=0.9\linewidth]{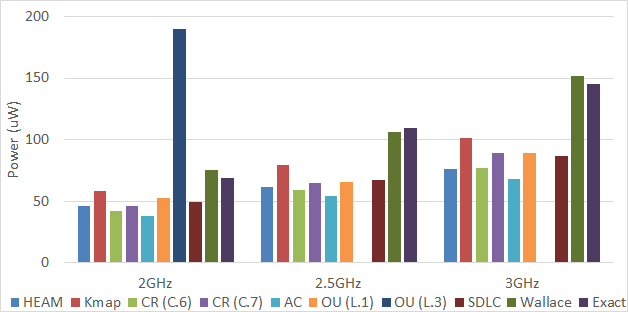}
\end{minipage}
}
\caption{The area and power consumption comparisons of the unsigned 8$\times$8 multipliers at fixed frequency. }
\label{Fig:compare_cost}
\end{figure*}

The proposed method focuses on partial product reduction to obtain an approximate multiplier with low hardware cost. Based on this idea, we use logic operations to compress the partial products of a multiplier. Fig.~\ref{Fig:opti_method} illustrates an example of how the partial products of an unsigned 4$\times$4 multiplier are compressed by logic operations. In Fig.~\ref{Fig:opti_method_a}, each \textbf{element} in the shape of a square or circle represents a bit and each row of elements is a \textbf{partial product}. Each element with an operator symbol inside in Fig.~\ref{Fig:opti_method_b} represents a \textbf{compressed term} generated by the corresponding logic operation on the bits represented by the same color and shape in Fig.~\ref{Fig:opti_method_a}. We use the symbols $\&$, $|$, and $\oplus$ to denote AND, OR, and XOR operations, respectively. The elements above the blue line constitute the \textbf{partial product bit-matrix} of the multiplier, where each row of the compressed terms forms a compressed partial product. The output of the 4$\times$4 approximate multiplier is obtained by the following steps: 

\begin{itemize}

\item Firstly, the partial products are divided into 10 \textbf{groups} as shown in Fig.~\ref{Fig:opti_method_a}, each of which includes one or two bit(s). The elements in the same group have the same shape and color. 

\item Secondly, AND, OR, and XOR operations are used to compress the groups into the compressed terms, which can be placed on any columns in the partial product bit-matrix. There are some elements without any symbol in Fig.~\ref{Fig:opti_method_b} because we do not apply any logic operation to the groups that contain only one bit. 

\item Finally, all partial products are summed as the output of the approximate multiplier. 

\end{itemize}

We denote a possible compressed term as $T_i(x, y) \in \{0, 1\}$, which is related to $L_i(x, y)$ in \eqref{Eqn:Formulation}. If a compressed term appears on the $c$-th column from the right in the compressed partial product bit-matrix, the corresponding $L_i(x, y)$ is $2^{c-1} \times T_i(x, y)$. Each weight $\mathbf{\theta}_i \in \{0, 1\}$ indicates whether the corresponding compressed term $T_i(x, y)$ is in the compressed partial product bit-matrix. Note that a possible compressed term has three attributes: the logic operation, the group, and the column. Thus, for a partial product bit-matrix with $A$ two-bits groups and $B$ columns, the number of basic operations $K$ is $3 \times A \times B$ in \eqref{Eqn:Formulation}. 

In a multiplier with a bit-width of more than 4, we may not compress all partial products. For example, we only compress the first four partial products in an 8$\times$8 multiplier. In this case, the approximate multiplication in our method is: 
\begin{equation}
\label{Eqn:f}
f(x, y \vert \mathbf{\theta}) = u(x, y) + \sum\limits_{i=0}^{Z-1} \theta_i L_i(x, y) 
\end{equation}
where $u(x, y)$ is the sum of the uncompressed partial products. The number of possible compressed terms $Z$ is less than $K$ in \eqref{Eqn:Formulation} because only part of the partial products are compressed.  

To reduce the number of compressed terms, we add a penalty $R(\mathbf{\theta}) = \lambda_1 \sum_{k=0}^{Z-1} \theta_k$ to the optimization objective, where the constant $\lambda_1$ controls the strength of the penalty. The final optimization objective is:
\begin{equation}
\label{Eqn:Min}
\begin{aligned}
\mathop{min}\limits_\mathbf{\theta} \Big( \sum_{i=0}^{N-1} \sum_{j=0}^{M-1} (x_iy_j - f(x_i, y_j \vert \mathbf{\theta}))^2  p(x_i, y_j) + R(\mathbf{\theta}) \Big)
\end{aligned}
\end{equation}
In this paper, the approximate multipliers are obtained by minimizing \eqref{Eqn:Min} with MATLAB Mixed Integer Genetic Algorithm. 


For multipliers with high bit widths, directly computing \eqref{Eqn:FixedPointError} may cost too much time and space. For example, we need to sum up $2^{16} \times 2^{16}$ values to compute \eqref{Eqn:FixedPointError} for a 16$\times$16 multiplier, which significantly slows down the optimization process. To solve this problem, we can randomly sample the operand values to estimate the average error. 


\subsection{The Generated Multiplier for DNNs}




We follow a widely-used quantization scheme \cite{DNN:Quantization} to train and evaluate DNNs with 8-bit integers in all experiments. The first four partial products of an 8$\times$8 multiplier are compressed in the optimization procedure as shown in Fig.~\ref{Fig:partial_products}. 

To obtain an approximate multiplier for DNNs, we analyze a quantized LeNet \cite{DNN:MNIST} trained on MNIST dataset and get the probability distributions of the inputs and weights of all layers. As examples of operand distributions, the histograms of the inputs and weights of its first fully-connected layer ($FC1$) are shown in Fig.~\ref{Fig:FC1_data} and Fig.~\ref{Fig:FC1_weight}, respectively. The inputs and weights are concentrated around 0 and 128, respectively. 

The structure of the optimized multiplier is presented in Fig.~\ref{Fig:appro_mul}. The size of the partial product bit-matrix is significantly reduced with negligible accuracy losses. 



\subsection{Evaluation Toolbox}

We implement a toolbox named \textit{ApproxFlow} to evaluate the DNN accuracy with approximate multiplication. In \textit{ApproxFlow}, each approximate multiplier is described by a look-up table. A DNN is represented by a directional acyclic graph (DAG), where each vertex denotes a DNN layer and the edges indicate the data flow. When a vertex in the DAG is executed, its dependencies will be executed automatically. We decouple \textit{ApproxFlow} from specific DNN toolboxes, unlike TFApprox \cite{Approximate:TFApprox} and ProxSim \cite{Approximate:ProxSim} which are based on Tensorflow \cite{DNN:tensorflow}.  




\section{Experiments}  \label{Sec:Exp}


\subsection{Comparison of Approximate Multipliers}

\begin{table*}[tbp]
\caption{Comparison of the Hardware Costs and Accuracies of the Multipliers on MNIST dataset}
\begin{center}
\begin{tabular}{|c|c|c|c|c|c|c|c|c||c|c|}
\hline
\textbf{Metric} & \textbf{HEAM} & \textbf{KMap} & \textbf{CR (C.6)} & \textbf{CR (C.7)} & \textbf{AC} & \textbf{OU (L.1)} & \textbf{OU (L.3)} & \textbf{SDLC} & \textbf{Exact} & \textbf{Wallace} \\
\hline
Area@$3GHz$ ($\mu m^2$) & 17.53 & 20.81 & 21.26 & 24.41 & 17.98 & 24.90 & - & \textbf{16.13} & 31.84 & 42.98 \\
\hline
Power@$3GHz$ ($\mu W$) & 76.20 & 101.67 & 77.15 & 89.38 & \textbf{68.44} & 89.51 & - & 86.55 & 145.49 & 151.94 \\
\hline
Delay ($ps$) & 247.94 & 256.94 & 226.98 & 239.88 & 227.90 & 244.91 & 463.96 & \textbf{225.00} & 297.93 & 278.99 \\
\hline
LeNet Accuracy (\%)& \textbf{99.34} & 98.33 & 57.20 & 95.60 & 21.14 & 10.32 & 95.70 & 95.51 & 99.40 & 99.40 \\
\hline
\end{tabular}
\label{Tab:Comparison1}
\end{center}
\end{table*}

\begin{table*}[tbp]
\caption{Comparison of the Accuracies on FashionMNIST, CIFAR-10, and CORA Dataset (\%)}
\begin{center}
\begin{tabular}{|c|c|c|c|c|c|c|c|c||c|c|}
\hline
\textbf{Dataset} & \textbf{HEAM} & \textbf{KMap} & \textbf{CR (C.6)} & \textbf{CR (C.7)} & \textbf{AC} & \textbf{OU (L.1)} & \textbf{OU (L.3)} & \textbf{SDLC} & \textbf{Exact} & \textbf{Wallace} \\
\hline
FashionMNIST & \textbf{91.01} & 81.66 & 20.88 & 78.34 & 17.53 & 15.16 & 49.95 & 75.89 & 91.26 & 91.26 \\
\hline
CIFAR-10      & \textbf{88.39} & 38.15 & 10.00 & 10.00 & 10.00 & 10.01 & 61.46 & 10.00 & 88.47 & 88.47 \\
\hline
CORA         & \textbf{80.24} & 79.58 & 76.4 & 79.39 & 70.94 & 15.44 & 30.21 & 78.80 &  80.24 &  80.24 \\
\hline
\end{tabular}
\label{Tab:Comparison2}
\end{center}
\end{table*}

\begin{table*}[tbp]
\caption{Comparison of DNN Accelrator Modules with the Approximate Multipliers}
\begin{center}
\begin{tabular}{|c|c|c|c|c|c|c|c|c||c|c|}
\hline
\textbf{Module} & \textbf{Metric} & \textbf{HEAM} & \textbf{KMap} & \textbf{CR (C.6)} & \textbf{CR (C.7)} & \textbf{AC} & \textbf{OU (L.1)} & \textbf{SDLC} & \textbf{Exact} & \textbf{Wallace} \\
\hline
TASU & Area@$1.5GHz$ ($\mu m^2$) & \textbf{84726.74} & 90209.57 & 90568.82 & 91023.57 & 87283.42 & 88737.03 & 86294.51 & 90011.99 & 98573.89 \\
\cline{2-11}
\cite{Accelerator:JiaoLi} & Power@$1.5GHz$ ($m W$) & \textbf{151.87} & 157.60 & 157.50 & 158.49 & 156.63 & 154.27 & 157.28 & 157.37 & 158.17 \\
\hline
\hline

SC  & Area@$2.5GHz$ ($\mu m^2$) & \textbf{3462.40} & 3709.50 & 3698.00 & 3683.87 & 3581.05 & 3546.40 & 3569.24 & 4258.61 & 4831.81 \\
\cline{2-11}
\cite{Accelerator:SystolicCube} & Power@$2.5GHz$ ($m W$) & 17.06 & 17.84 & 17.69 & 17.76 & 17.60 & \textbf{16.76} & 17.75 & 18.37 & 18.26 \\
\hline
\hline

SA & Area@$2GHz$ ($\mu m^2$) & 25508.38 & - & 24473.36 & 24858.45 & 22839.51 & - & \textbf{22105.41} & 24926.97 & - \\
\cline{2-11}
\cite{Accelerator:TPU} & Power@$2GHz$ ($m W$) & \textbf{74.76} & - & 82.38 & 82.50 & 81.52 & - & 81.56 & 83.06 & - \\
\hline
\end{tabular}
\label{Tab:Comparison3}
\end{center}
\end{table*}

In this section, we compare our multiplier with several reproduced 8$\times$8 approximate multipliers from recent works, including KMap \cite{Approximate:KMap}, Configurable-Recovery (CR) \cite{Approximate:Configurable}, Approximate-Compressor (AC) \cite{Approximate:Compressor}, Optimal-Unbiased (OU) \cite{Approximate:Optimal}, and SDLC \cite{Approximate:BitMatrix} multipliers. We reproduce two CR multipliers with 6-bit (C.6) and 7-bit (C.7) error recovery units. Two OU multipliers are implemented with level-1 (L.1) and level-3 (L.3) structures. Note that we implement OU multipliers by applying the optimization method in \cite{Approximate:Optimal} to 8$\times$8 integer multipliers to provide a fair comparison with others. In addition to the approximate multipliers, we use the standard multiplier synthesized by Synopsys Design Compiler (DC) 2016 as the exact multiplier for comparison. A Wallace Tree multiplier \cite{Approximate:WallaceTree} is also implemented. The multipliers are evaluated with the following metrics: (1) The areas and power consumptions at fixed frequencies. (2) The minimum critical-path delays of the multipliers. (3) The LeNet accuracies on MNIST dataset. 






To evaluate the multipliers, we implement them in Verilog, and synthesize the circuits in DC with Arizona State Predictive PDK (ASAP) 7$nm$ process libraries \cite{VLSI:ASAP7}. The areas, power consumptions, and delays are obtained from DC. Fig.~\ref{Fig:compare_cost} presents the comparison of areas and power consumptions among the multipliers, where HEAM shows its competitive hardware efficiency. Some data are absent because the corresponding multipliers fail to meet the timing constraints. In Table~\ref{Tab:Comparison1}, we compare the hardware costs and DNN accuracies on MNIST dataset of the multipliers. The areas and power consumptions presented in Table~\ref{Tab:Comparison1} are obtained at $3GHz$. 

As shown in Table~\ref{Tab:Comparison1}, HEAM achieves the best accuracy among the approximate multipliers. Compared with KMap, HEAM has 15.76\% smaller area, 25.05\% less power consumption, 3.50\% shorter delay, and 1.01\% higher accuracy. Furthermore, HEAM attains 44.94\% smaller area, 47.63\% less power consumption, and 16.78\% shorter delay than the exact multiplier with negligible accuracy loss. In terms of hardware cost, HEAM has slightly larger area than SDLC and higher power consumption than AC, but SDLC and AC suffer from nonnegligible accuracy loss. In conclusion, HEAM can achieve the highest accuracy among the tested approximate multipliers with small area and low power consumption. 

\subsection{Accuracy Comparison on Different Datasets}

In addition to MNIST, we carry out an experiment on FashionMNIST \cite{DNN:FashionMNIST} with LeNet structure and CIFAR-10 \cite{DNN:CIFAR} with AlexNet \cite{DNN:alexnet} structure. Moreover, the multipliers are evaluated in a two-layer graph convolutional network (GCN) on CORA dataset \cite{DNN:GCN}. We use the multiplier generated from the LeNet on MNIST dataset in this experiment rather than design a multiplier for each dataset. 

Table~\ref{Tab:Comparison2} presents the accuracies of the neural networks with the multipliers. HEAM achieves the highest accuracies among the approximate multipliers, obtaining 9.35\%, 50.24\%, and 0.66\% higher accuracies than KMap on FashionMNIST, CIFAR-10, and CORA datasets, respectively. KMap surpasses other approximate multipliers except HEAM on MNIST, FashionMNIST, and CORA datasets. Although OU (L.3) has better accuracy than KMap on CIFAR-10 dataset, it suffers from large hardware cost as shown in Fig.~\ref{Fig:compare_cost}. Thus, KMap can be regarded as the best reproduced approximate multiplier. The experiment on these datasets further proves that the proposed optimization method can obtain negligible accuracy loss. 

\subsection{Comparison of DNN Accelerator Modules with Approximate Multipliers}



In this experiment, we compare the multipliers by applying them to several DNN accelerator modules. TASU \cite{Accelerator:JiaoLi} is a DNN accelerator for DoReFa-Net \cite{DNN:DoReFaNet}. We synthesize its processing block for the first convolution layer on DC to compare the multipliers. We also reproduce Systolic Cube (SC) \cite{Accelerator:SystolicCube}, an efficient acceleration module for the convolution operations in DNNs. Systolic Array (SA) is a popular accelerator module adopted by Google Tensor Processing Unit (TPU) \cite{Accelerator:TPU}. We implement a 16$\times$16 SA in our experiment as well. 

Table~\ref{Tab:Comparison3} presents the comparison of the accelerator modules with the multipliers on 7$nm$, showing their areas and power consumptions at fixed frequencies. Some results are not shown in the table because the corresponding modules fail to meet the timing constraints. Due to the space limit, we do not show the results of OU (L.3), whose hardware cost is obviously higher than others as shown in Fig.~\ref{Fig:compare_cost}. HEAM attains the lowest areas and power consumptions in the TASUs and obtains competitive hardware costs in other modules. The modules with HEAM obtain up to 6.66\% smaller area and 4.43\% less power consumption than the modules with KMap. Moreover, the modules with HEAM achieve up to 18.70\% area reduction and 9.99\% power saving compared to the modules with the exact multiplier. This experiment indicates that HEAM can effectively reduce the hardware cost of DNN accelerators. 


\section{Conclusion}

We propose an optimization method to design a high-efficiency approximate multiplier, which minimizes the average error according to the probability distributions of operands in the target application. The generated approximate multiplier can achieve negligible accuracy loss with small area, low power consumption, and short delay. Our multiplier and \textit{ApproxFlow} are available at \textit{https://github.com/FDU-ME-ARC/HEAM} and  \textit{https://github.com/FDU-ME-ARC/ApproxFlow}.

\bibliography{IEEEexample}{}
\bibliographystyle{IEEEtran}

\end{document}